\documentclass{article}
\usepackage{spconf,amsmath,graphicx}
\usepackage{multirow,makecell,adjustbox,xcolor}

\usepackage{amsfonts}



\makeatletter
\def\section{\@startsection{section}{1}{\z@}%
  {-1.5ex \@plus -0.5ex \@minus -.2ex}%
  {0.8ex \@plus .2ex}%
  {\large\bfseries}}
\def\subsection{\@startsection{subsection}{2}{\z@}%
  {-1.2ex \@plus -0.5ex \@minus -.2ex}%
  {0.6ex \@plus .2ex}%
  {\normalsize\bfseries}}
\makeatother

\title{MTS-CSNet: Multiscale Tensor Factorization for Deep Compressive Sensing on RGB Images}

\name{
Mehmet Yama\c{c}$^{1}$ \quad
Lei Xu$^{1}$ \quad
Serkan Kiranyaz$^{2}$ \quad
Moncef Gabbouj$^{1}$
}

\address{
$^{1}$Faculty of Information Technology and Communication Sciences, Tampere University, Finland\\
$^{2}$Department of Electrical Engineering, Qatar University, Doha, Qatar\\
\{mehmet.yamac, lei.xu, moncef.gabbouj\}@tuni.fi \quad serkan.kiranyaz@qu.edu.qa
}

\begin{document}
%
\maketitle
\begin{abstract}
Deep learning–based compressive sensing (CS) methods typically learn sampling operators using convolutional or block-wise fully connected layers, which limit receptive fields and scale poorly for high-dimensional data. We propose MTS-CSNet, a CS framework based on Multiscale Tensor Summation (MTS) factorization, a structured operator for efficient multidimensional signal processing. MTS performs mode-wise linear transformations with multiscale summation, enabling large receptive fields and effective modeling of cross-dimensional correlations. In MTS-CSNet, MTS is first used as a learnable CS operator that performs linear dimensionality reduction in tensor space, with its adjoint defining the initial back-projection, and is then applied in the reconstruction stage to directly refine this estimate. This results in a simple feed-forward architecture without iterative or proximal optimization, while remaining parameter and computation efficient. Experiments on standard CS benchmarks show that MTS-CSNet achieves state-of-the-art reconstruction performance on RGB images, with notable PSNR gains and faster inference, even compared to recent diffusion-based CS methods, while using a significantly more compact feed-forward architecture.
\end{abstract}
\section{Introduction}
\label{sec:intro}

Compressive sensing (CS) provides a principled framework for acquiring high-dimensional signals using substantially fewer measurements than required by conventional sampling strategies, by exploiting underlying structure such as sparsity or low-dimensional representations~\cite{CS1}. Owing to this capability, CS has found widespread use in applications including magnetic resonance imaging (MRI)~\cite{MR}, radar and wireless monitoring~\cite{radar1}, biomedical signal acquisition~\cite{ecg}, as well as privacy-preserving data acquisition and encryption~\cite{CSinEnc,yamacEn}. Despite its strong theoretical foundations, the practical adoption of classical CS methods is often hindered by the reconstruction stage, which typically relies on iterative optimization procedures. Traditional approaches based on convex relaxation and $\ell_1$-minimization~\cite{BP,BPDN} provide theoretical recovery guarantees, yet they incur high computational cost and exhibit sensitivity to measurement noise, model mismatch, or deviations from ideal sparsity assumptions. Although a variety of accelerated solvers have been proposed~\cite{TVAL3,figueiredo2007gradient}, their inherently iterative nature limits scalability to large-scale, multidimensional data such as images and videos, motivating alternative formulations that enable more efficient and robust reconstruction.

With the advent of deep learning, data-driven recovery frameworks have fundamentally reshaped the CS landscape. A first category of learning-based CS methods consists of \emph{feed-forward reconstruction networks}, which employ neural networks solely to map compressed measurements to reconstructed signals while using fixed, typically random, sensing operators~\cite{reconnet, CSNET}. These approaches offer fast inference and simple architectures, but their reconstruction performance is often limited by the expressiveness of purely non-iterative mappings. A second line of work builds upon classical optimization algorithms through \emph{deep unfolding} or unrolling strategies, where iterative CS solvers are parameterized and truncated into trainable networks~\cite{ISTA-Net}. By embedding domain knowledge from iterative shrinkage or thresholding schemes, these methods achieve improved reconstruction accuracy and stability at the cost of increased architectural complexity and iterative computation. More recently, \emph{diffusion-based CS models}~\cite{Rout2023SolvingLI, chen2025invertible} have emerged as a powerful alternative, leveraging generative denoising processes to progressively refine reconstructions from compressed measurements. While diffusion-based approaches can achieve impressive reconstruction quality, they typically require many iterative sampling steps, resulting in substantial computational overhead and slow inference. 

Across these categories, a common limitation is that sensing operators are typically implemented using convolutional or fully connected layers applied to local image blocks. While computationally convenient, such block-wise designs restrict the ability to model global signal dependencies and often introduce blocking artifacts or loss of fine details, especially at low measurement rates. To address these limitations, \emph{Generalized Tensor Summation (GTS)}~\cite{GTS} was proposed as a structured sensing operator that performs multilinear transformations directly in tensor space, enabling global interactions without resorting to block-wise approximations. By factorizing the sensing operation across tensor modes, GTS provides a principled alternative to conventional block-based CS operators, demonstrating improved reconstruction quality and parameter efficiency in compressive sensing settings.

Recent advances in multiscale tensor-based layers~\cite{yamacc2025multiscale,Xu2025MultiScaleTS} have introduced a new class of generalized linear operators that serve as efficient alternatives to conventional MLP and convolutional layers. Building on this line of work, we employ the Multiscale Tensor Summation (MTS) operator as a learnable CS operator that performs linear dimensionality reduction directly in tensor space, together with its adjoint for initial signal recovery. The adjoint output is then refined using a small number of MTSNet blocks~\cite{yamacc2025multiscale}, resulting in a simple feed-forward reconstruction architecture without iterative optimization or deep unfolding. Based on this design, we introduce the \emph{Multiscale Tensor Summation for Compressive Sensing (MTS-CS)} framework, a unified sensing–reconstruction system that operates entirely in tensor space. Unlike block-based or separable CS formulations, MTS-CS leverages a multiscale tensorial factorization to jointly model spatial and channel-wise correlations across scales using a single learnable linear operator. This unified formulation enables parameter-efficient and high-quality reconstruction, particularly at low measurement ratios, and establishes multiscale tensor operators as an effective foundation for compressive sensing of multidimensional data. Despite its non-iterative, feed-forward design, MTS-CSNet consistently surpasses both conventional deep CS networks and recent diffusion-based models in reconstruction quality, while requiring significantly lower computational complexity.




This paper makes the following contributions: (i) We introduce MTS-CSNet, which employs the Multiscale Tensor Summation (MTS) operator as a learnable compressive sensing operator, enabling structured, multiscale dimensionality reduction directly in tensor space. (ii) We propose a nonlinear adjoint back-projection by incorporating an MHG activation prior to the adjoint MTS operator, improving the expressiveness of the initial reconstruction compared to conventional linear adjoints. (iii) We develop a fully end-to-end, non-iterative CS framework that achieves reconstruction quality on par with or exceeding state-of-the-art diffusion-based CS methods, while being orders of magnitude faster and significantly more computationally efficient.

The remainder of this paper is organized as follows. Section~\ref{sec:meth} presents the proposed MTS-CS formulation and network architecture. Section~\ref{sec:exp} reports experimental results and comparisons with state-of-the-art methods. Finally, Section~\ref{sec:con} concludes the paper and discusses future directions.


\section{Preliminaries and Background}
\label{sec:prelim}
\noindent\textbf{Classical Compressive Sensing:}
Compressive sensing~\cite{CS1,CS2} provides a theoretical framework for acquiring and reconstructing signals from a reduced number of linear measurements by exploiting low-dimensional structure, such as sparsity. The measurement process is typically modeled as
\begin{equation}
    \mathbf{y} = \mathbf{\Psi}\mathbf{s},
\end{equation}
where $\mathbf{s} \in \mathbb{R}^N$ denotes the signal of interest and $\mathbf{\Psi} \in \mathbb{R}^{m \times N}$ is the sensing operator with $m \ll N$. Classical CS theory establishes conditions under which stable recovery is possible, relying on properties such as the Restricted Isometry Property (RIP) and incoherence of the sensing matrix~\cite{candesRIP}. While convex optimization-based recovery methods~\cite{BP,BPDN} offer strong theoretical guarantees, their reliance on iterative solvers leads to high computational cost, limiting their practicality for large-scale and high-dimensional data such as images and videos.

\noindent\textbf{Separable and Tensor-Structured CS:}
To reduce the computational burden of large-scale sensing operators, separable and block-based CS formulations~\cite{separableCS,kronecker} approximate high-dimensional measurement matrices using Kronecker or tensor-product structures. In this setting, the sensing process can be expressed as
\begin{equation}
    \mathcal{Y} = \mathcal{S} \times_1 \mathbf{\Psi}_1 \times_2 \mathbf{\Psi}_2 \cdots \times_J \mathbf{\Psi}_J,
\end{equation}
where $\mathcal{S}$ denotes a multidimensional signal and $\mathbf{\Psi}_j$ represent mode-wise projection matrices applied along each tensor dimension. Such separable constructions significantly reduce memory requirements and computational cost by decomposing a large linear operator into a set of smaller factors. However, this structural constraint also restricts representational capacity, often increasing mutual coherence and leading to degraded reconstruction performance at low measurement ratios. 

\noindent\textbf{Generalized Tensor Summation Compressive Sensing (GTS-CS): }
\label{sec:GTS-CS}
The GTS framework~\cite{GTS} introduces a structured and learnable formulation for approximating non-separable CS operators using tensor summation. Unlike conventional separable or Kronecker-based CS schemes, where sub-matrices are typically fixed random or randomized projections~\cite{kronecker}, GTS parameterizes the sensing operator itself and enables end-to-end learning of the measurement process. By expressing a dense sensing matrix as a summation of multiple separable tensor contractions, GTS bridges the gap between fully unstructured random projections and computationally efficient but limited separable CS designs, while significantly enhancing representational capacity.

\begin{figure*}[t]
\begin{center}
    \includegraphics[width=0.86\linewidth]{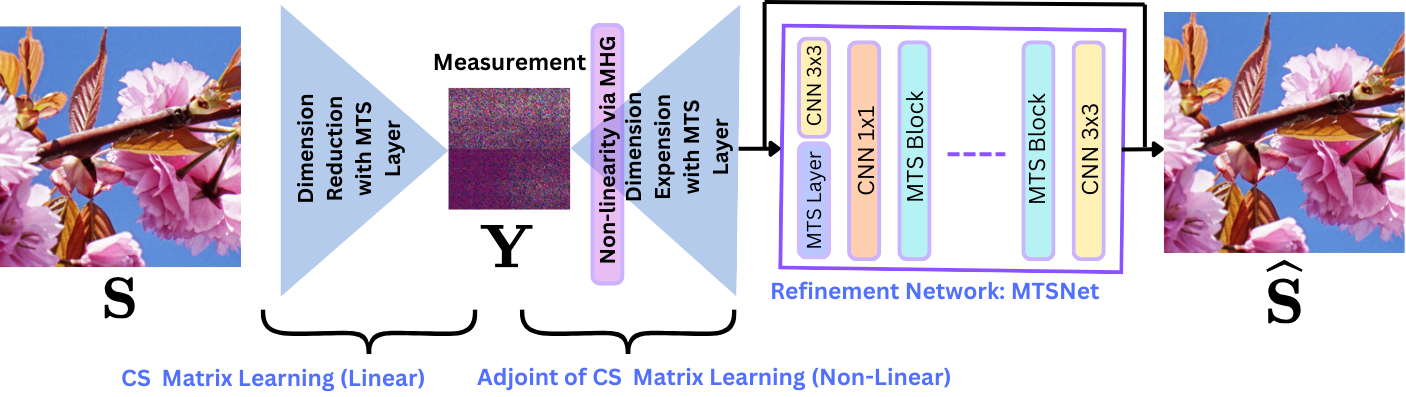}
\end{center}
\vspace{-0.7cm}
   \caption{Overall architecture of the proposed MTS-CSNet. An MTS-layer-based autoencoder performs compressive sensing and adjoint back-projection, followed by a lightweight MTSNet refinement network for feed-forward reconstruction.
}
\label{fig:overall}
\end{figure*}

Let $\mathcal{S} \in \mathbb{R}^{n_1 \times n_2 \times \dots \times n_J}$ denote a $J$-dimensional input signal. The GTS-CS measurement process is defined as
\begin{equation}
    \mathcal{Y} = \sum_{t=1}^{T} \mathcal{S} 
    \times_1 \mathbf{\Psi}_1^{(t)} 
    \times_2 \mathbf{\Psi}_2^{(t)} 
    \cdots 
    \times_J \mathbf{\Psi}_J^{(t)},
    \label{eq:gts_cs_tensor}
\end{equation}
where $\mathbf{\Psi}_j^{(t)} \in \mathbb{R}^{m_j \times n_j}$ denotes the learnable projection matrix along the $j$-th tensor mode at the $t$-th summation component, and $T$ controls the number of tensor summation terms. In vectorized form, this operation can be written as
\begin{equation}
    \mathbf{y} = 
    \sum_{t=1}^{T} 
    \left( 
    \mathbf{\Psi}_1^{(t)} 
    \otimes 
    \mathbf{\Psi}_2^{(t)} 
    \otimes 
    \cdots 
    \otimes 
    \mathbf{\Psi}_J^{(t)} 
    \right) 
    \mathbf{s} 
    = 
    \mathbf{P}\mathbf{s},
    \label{eq:gts_cs_vector}
\end{equation}
where $\mathbf{s} = \mathrm{vec}(\mathcal{S})$ and $\mathbf{P}$ denotes the effective GTS-CS sensing matrix. When $T = 1$, the formulation reduces to the classical separable Kronecker CS model~\cite{kronecker}. Increasing $T$ yields progressively richer non-separable representations, allowing the effective sensing matrix to approximate a fully dense projection while retaining a structured and efficient parameterization. By aggregating multiple separable components through summation, GTS substantially improves the expressive power of the sensing operator compared to single-factor separable designs, enabling improved modeling of global correlations. At the same time, it preserves the favorable incoherence properties of random projections and reduces the parameter complexity from $\mathcal{O}(mN)$ to $\mathcal{O}\!\left(T \sum_{j} m_j n_j\right)$, making it well suited for high-dimensional compressive sensing applications.

\vspace{-0.1cm}
\section{Proposed Solution}
\label{sec:meth}
\subsection{Multiscale Tensor Summation Layer}
\label{sec:mts_layer}
MTS operator~\cite{yamacc2025multiscale} is a structured linear transformation that operates directly in tensor space by aggregating multiple patch-wise multilinear projections across spatial scales. Given an input tensor $\mathcal{X}$, MTS applies mode-wise linear mappings to local patches extracted at a predefined set of window sizes $w = \{w_1, w_2, \dots, w_{SC}\}$, where $SC$ denotes the number of scales. The responses obtained at different scales are mapped back to the original spatial layout and summed, yielding a multiscale representation with enhanced expressive capacity. Formally, the MTS operation is defined as
\begin{equation}
\mathcal{Y} = \sum_{sc=1}^{SC} f^{-1}_{w_{sc}}
\left(
\sum_{t=1}^{T}
f_{w_{sc}}(\mathcal{X})
\times_1 \mathbf{A}_1^{(t,sc)}
\times_2 \cdots
\times_J \mathbf{A}_J^{(t,sc)}
\right),
\label{eq:mts}
\end{equation}
where $f_{w_{sc}}(\cdot)$ and $f^{-1}_{w_{sc}}(\cdot)$ denote the patch embedding and inverse embedding operators associated with window size $w_{sc}$. The matrices $\mathbf{A}_j^{(t,sc)}$ are learnable linear projections along the $j$-th tensor mode, and $T$ controls the number of separable tensor components aggregated at each scale.

From a linear operator viewpoint, MTS corresponds to a structured transformation composed of a sum of block-diagonal operators, each implementing a patch-wise tensor contraction. In vectorized form, the equivalent linear mapping can be written as
\begin{equation}
\mathbf{P}
=
\sum_{sc=1}^{SC}
\mathbf{I}_{B_{sc}}
\otimes
\sum_{t=1}^{T}
\left(
\mathbf{A}_1^{(t,sc)} \otimes
\mathbf{A}_2^{(t,sc)} \otimes \cdots \otimes
\mathbf{A}_J^{(t,sc)}
\right),
\label{eq:mts_matrix}
\end{equation}
where $\mathbf{I}_{B_{sc}}$ enforces the patch-wise structure. Although Eq.~\eqref{eq:mts_matrix} reveals the equivalence to a large linear sensing matrix, explicitly forming $\mathbf{P}$ is unnecessary in practice, as MTS can be efficiently implemented through tensor operations.

\begin{table*}[!htp]
\caption{Comparison with competing methods on $256 \times 256$ RGB images~\cite{Chen2024InvertibleDM, chen2024self}.}
\begin{center}
\begin{adjustbox}{width=1\textwidth,center}
\begin{tabular}{ c|c|cccccccc|cccc }
\hline
\hline
\multirow{1}{*}{Datasets}
&\multirow{1}{*}{CR}
    & \multicolumn{1}{c}{\thead{ DDRM \cite{Kawar2022DenoisingDR} \\ (NeurIPS 2022)}}
    & \multicolumn{1}{c}{\thead{$\Pi$GDM \cite{Song2023PseudoinverseGuidedDM} \\ (ICLR 2023)}}
    & \multicolumn{1}{c}{\thead{ DPS \cite{Chung2022DiffusionPS} \\ (ICLR 2023)}}
    & \multicolumn{1}{c}{\thead{DDNM \cite{Wang2022ZeroShotIR} \\ (ICLR 2023)}}
    &  \multicolumn{1}{c}{\thead{GDP\cite{Fei2023GenerativeDP} \\ (CVPR 2023)}}
   
    & \multicolumn{1}{c}{\thead{PSLD\cite{Rout2023SolvingLI} \\ (NeurIPS 2023)}}
    &  \multicolumn{1}{c}{\thead{SR3 \cite{Saharia2021ImageSV} \\ (TPAMI 2023)}} 
    & \multicolumn{1}{c|}{\thead{IDM \cite{chen2025invertible} \\ (TPAMI 2025)}}
    
        &  \multicolumn{1}{c}{\thead{MTS-CSNet \\ (T=12, NB=3)}}
        &  \multicolumn{1}{c}{\thead{MTS-CSNet \\ (T=24, NB=3)}}
         &  \multicolumn{1}{c}{\thead{MTS-CSNet \\ (T=48, NB=3)}}
         
\\
\hline
\hline   
\multirow{4}{*}{Urban100}& 10\% &19.16/0.4348& 20.09/0.5089& 17.12/0.3270&20.76/0.4682& 20.74/0.5075&19.43/0.4054&18.90/0.5023& \textcolor{red}{\underline{30.85}}/\textcolor{red}{\underline{0.8970}}&28.53/0.8773&28.91/0.8846& \textcolor{blue}{\underline{28.95}}/\textcolor{blue}{\underline{0.8854}}& \\
                         & 30\%&28.91/0.8518&26.70/0.7925&18.47/0.3891&28.76/0.8282&24.81/0.7086&22.42/0.6399&21.37/0.6428&35.78/0.9570 &\textcolor{blue}{\underline{36.50}}/\textcolor{blue}{\underline{0.9709}}& \textcolor{red}{\underline{36.83}}/\textcolor{red}{\underline{0.9719}}&35.14/0.9614&\\
                         & 50\%& 33.61/0.9365&29.75/0.8724&19.21/0.4308&32.86/0.9164&26.12/0.7711&22.78/0.6882&23.12/0.7233&39.16/0.9771&\textcolor{blue}{\underline{43.12}}/\textcolor{blue}{\underline{0.9918}}&40.89/0.9893& \textcolor{red}{\underline{45.70}}/\textcolor{red}{\underline{0.9942}}&\\
                     
                         & Avg. &27.38/0.7410 &25.51/0.7246&18.27/0.3823&27.46/0.7376&23.89/0.6624& 21.54/0.5778&21.13/0.6228&35.26/0.9437&\textcolor{blue}{\underline{36.05}}/\textcolor{blue}{\underline{0.9467}}& 35.54/0.9486&\textcolor{red}{\underline{36.60}}/\textcolor{red}{\underline{0.9470}} \\
\hline
\multirow{4}{*}{DIV2K}& 10\% &20.91/0.4323&22.46/0.5551&19.47/0.4222&22.18/0.4383&25.17/0.6334&21.30/0.4427&20.20/0.4794&31.22/0.8581&32.91/0.9230 &\textcolor{blue}{\underline{33.15}}/\textcolor{blue}{\underline{0.9255}}&\textcolor{red}{\underline{33.22}}/\textcolor{red}{\underline{0.9265}}& \\
                         & 30\%&28.91/0.7852&28.01/0.7714&20.78/0.4652&28.50/0.7422&27.75/0.7664&23.87/0.6473&22.07/0.5689&36.83/0.9482&\textcolor{blue}{\underline{40.85}}/\textcolor{blue}{\underline{0.9833}}&\textcolor{red}{\underline{41.22}}/\textcolor{red}{\underline{0.9837}}&39.85/0.9775 \\
                         & 50\%&33.68/0.9057&30.79/0.8497&21.37/0.4884&32.74/0.8726&28.47/0.8029&24.41/0.7124&23.63/0.6518&40.81/0.9755&\textcolor{blue}{\underline{47.79}}/\textcolor{blue}{\underline{0.9956}}&46.07/0.9948&\textcolor{red}{\underline{49.63}}/\textcolor{red}{\underline{0.9970}}&  \\
                    
                         & Avg. & 27.83/0.7077&27.09/0.7254&20.54/0.4586&27.81/0.6844&27.13/0.7342&23.19/0.6008&21.97/0.5667&36.29/0.9273&\textcolor{blue}{\underline{40.52}}/\textcolor{red}{\underline{0.9673}}&40.15/0.9680&\textcolor{red}{\underline{40.90}}/\textcolor{blue}{\underline{0.9670}} \\
\hline

\end{tabular}
\end{adjustbox}
\end{center}
\vspace{-0.5cm}
\label{tab:res1}
\end{table*}

\begin{figure*}[t]
\begin{center}
    \includegraphics[width=1.0\linewidth]{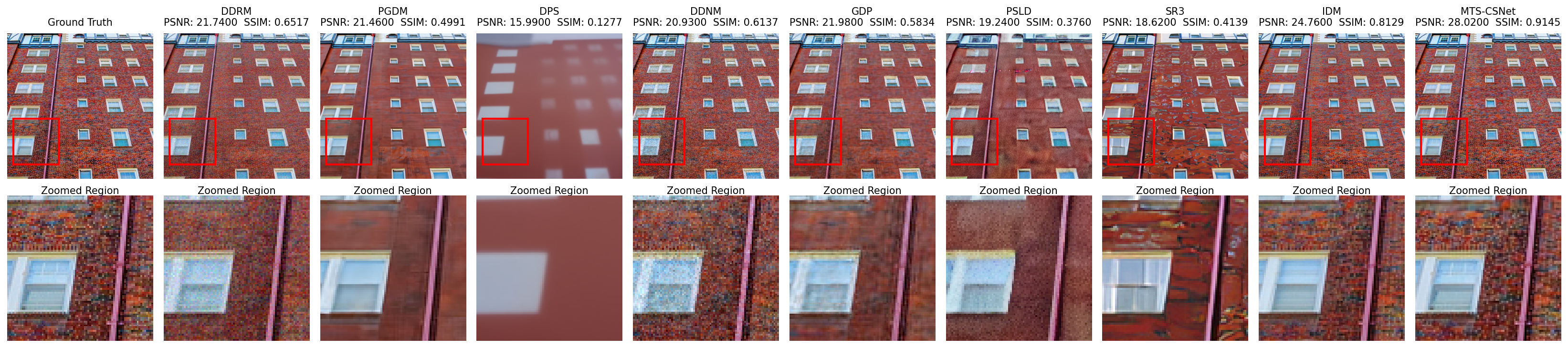}
\end{center}
   \vspace{-0.6cm}
   \caption{Visual comparison of CS reconstruction results on the ``img034'' image from Urban100 at sampling rate 0.3. }
\label{fig:example1}
\end{figure*}

\begin{figure*}[htbp]
\begin{center}
    \includegraphics[width=1.0\linewidth]{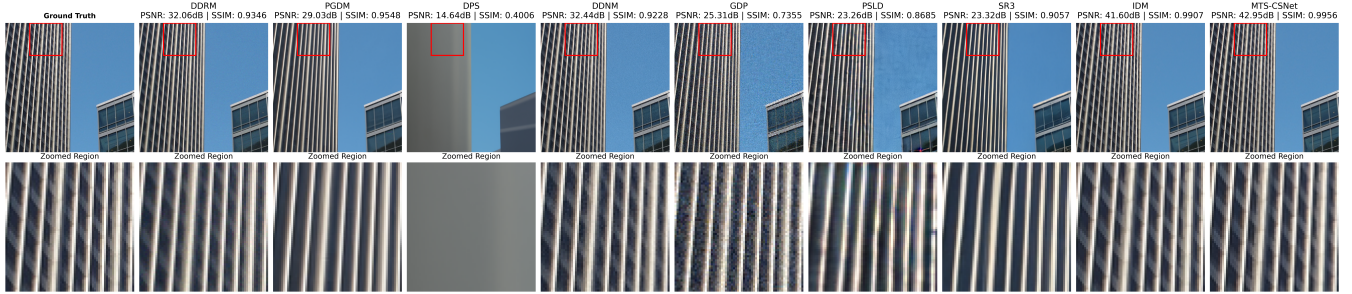}
\end{center}
   \vspace{-0.6cm}
   \caption{Comparison of CS recovery results among various competing methods on ``img096'' image from Urban100 at sampling rate 0.3.}
\label{fig:example2}
\end{figure*}

\subsection{MTS-Based Dimensional Reduction}
\label{sec:mts_forward}
In the proposed MTS-CSNet, the MTS layer is first instantiated as a compressive sensing operator. Given an input image tensor $\mathcal{S}$, Eq.~\eqref{eq:mts} is used to produce a lower-dimensional measurement tensor $\mathbf{Y}$ by selecting the output dimensions of the projection matrices $\{\mathbf{A}_j^{(t,sc)}\}$ according to the desired compression ratio. In this setting, the learnable projection matrices constitute the parameters of the sensing operator, enabling data-driven optimization of the measurement process while preserving a structured, multiscale formulation.

\subsection{Adjoint MTS and Proxy Reconstruction}
\label{sec:mts_adjoint}
To obtain an initial reconstruction from the compressed measurements, MTS-CSNet employs the adjoint of the MTS operator as a learnable back-projection. Unlike most existing CS formulations that rely on purely linear adjoint operators, we introduce a lightweight nonlinearity via the MHG operator~\cite{yamacc2025multiscale} prior to the adjoint mapping, enhancing representational flexibility without iterative optimization. Given the measurement tensor $\mathbf{Y}$, we first apply the MHG nonlinearity~\cite{yamacc2025multiscale}, after which the adjoint MTS applies the transpose of the mode-wise projection matrices in Eq.~\eqref{eq:mts} and performs inverse patch embedding to produce a proxy reconstruction $\tilde{\mathcal{S}}$. This proxy image serves as a coarse yet informative estimate of the original signal, capturing global spatial and cross-channel correlations.
\subsection{MTSNet-Based Refinement Network}
\label{sec:refine}

Following the adjoint MTS back-projection, the coarse reconstruction is further improved using an MTSNet-based refinement network~\cite{yamacc2025multiscale}. The refinement network consists of a small number of MTSBlocks, each composed of stacked MTS layers with shared multiscale tensor summation principles. Owing to its lightweight design, the refinement stage enhances reconstruction quality while maintaining low computational complexity. The overall architecture of the proposed MTS-CSNet, including the MTS-layer-based sensing operator, its adjoint back-projection, and the MTSNet refinement network, is illustrated in Fig.~\ref{fig:overall}.

\section{Experiments}
\label{sec:exp}

\subsection{Experimental Settings}
\textbf{Datasets.}
Following prior works~\cite{Chen2024InvertibleDM, chen2024self}, we evaluate MTS-CSNet on cropped RGB images from Urban100~\cite{Huang2015SingleIS} and DIV2K~\cite{Agustsson2017NTIRE2C}, using $256 \times 256$ patches. Additional evaluations are conducted on Urban100, Set5~\cite{Bevilacqua2012LowComplexitySS}, and Set14~\cite{Zeyde2010OnSI}. During training, images are randomly cropped to $256 \times 256$, and MTS window sizes are set to $[20, 40, 80, 160]$.

\noindent
\textbf{Implementation Details.}
All models are implemented in PyTorch and trained on an NVIDIA GPU cluster for 30k iterations with a batch size of 2. Optimization is performed using Adam~\cite{Kingma2014AdamAM} with a learning rate of $5\times10^{-4}$ and a cosine restart scheduler~\cite{Loshchilov2016SGDRSG}. Reconstruction quality is evaluated using PSNR and SSIM~\cite{Wang2004ImageQA}.

\noindent
\textbf{Hyperparameters.}
The MTS-layer-based autoencoder uses 3 input and output channels. For a given compression ratio ($CR$), the output window size is set as $\hat{w}_i = \sqrt{w_i \cdot CR}$. The refinement network consists of $NB$ MTS blocks, each containing four MTS layers, with window scales $[8, 16, 32, 64]$ and $T=3$. Network scalability is examined by varying $T$, $NB$, and window scales in the autoencoder.

\subsection{Quantitative and Visual Results}
For Urban100 and DIV2K, we train three MTS-CSNet variants with $(T, NB) \in \{(12,3), (24,3), (48,3)\}$ following~\cite{chen2025invertible}. We compare against eight state-of-the-art CS methods: DDRM~\cite{Kawar2022DenoisingDR}, $\Pi$GDM~\cite{Song2023PseudoinverseGuidedDM}, DPS~\cite{Chung2022DiffusionPS}, DDNM~\cite{Wang2022ZeroShotIR}, GDP~\cite{Fei2023GenerativeDP}, PSLD~\cite{Rout2023SolvingLI}, SR3~\cite{Saharia2021ImageSV}, and IDM~\cite{chen2025invertible}, at sampling rates $CR \in \{10\%, 30\%, 50\%\}$. Quantitative results are reported in Table~\ref{tab:res1}, where best and second-best performances are highlighted. MTS-CSNet consistently achieves state-of-the-art results, with the $(T=48, NB=3)$ model outperforming IDM by 1.34~dB and 4.61~dB in average PSNR on Urban100 and DIV2K, respectively, and improving SSIM by 0.0043 and 0.0387. In particular, MTS-CSNet attains the best or second-best performance at $30\%$ and $50\%$ sampling rates across both datasets. Representative visual comparisons are shown in Figs.~\ref{fig:example1} and~\ref{fig:example2}. MTS-CSNet recovers fine structures and sharp edges, while competing methods exhibit noticeable blurring and artifacts.

\subsection{Complexity Analysis}
Table~\ref{tab:complex} compares model complexity in terms of parameters, memory usage, GFLOPs, and inference time at a sampling ratio of $10\%$ on DIV2K. As shown, MTS-CSNet achieves substantially lower computational cost and memory consumption than competing methods, while maintaining superior reconstruction performance, demonstrating the efficiency of its lightweight feed-forward design.

\begin{table}[t]
\caption{Computational complexity comparison in terms of PSNR/SSIM, model size, memory, GFLOPs, and inference time on DIV2K at a CS ratio of 0.1.
}
\vspace{0.2cm}
\begin{adjustbox}{width=0.45\textwidth,center}
 \begin{tabular}{c|c |cccc} 
 \hline
 \hline
 Networks    & DIV2K &Params &Memory &  GFLOPs & Time\\  
 \hline
 \hline
  \thead{ ($T=12$, $NB=3$)} &32.91/0.9230 &1.69M& 0.0063GB&154.77&0.0066s\\ 
  \thead{ ($T=24$, $NB=3$)} &33.15/0.9255&2.27M& 0.0084GB&161.36&0.0066s\\ 
  \thead{ ($T=48$, $NB=3$)} &33.22/0.9265&3.41M&0.0127GB& 174.56&0.0067s\\ 
  \hline
  \hline
  \thead{SR3} & 20.20/0.4794 & -&0.4GB&-&35.62s \\ 
  \thead{PSLD} &21.30/0.4427 & -&4.3GB&-&233.10s \\ 
  \thead{DDNM} &22.18/0.4383  & -&2.1GB&-&21.35s \\ 
  \thead{IDM} & 31.22/0.8581 & -&0.4GB&-&0.63s \\ 
 \hline
 \hline
 \end{tabular}
 \end{adjustbox}
\label{tab:complex}
\end{table}

\vspace{-0.3cm}
\begin{table}[t]
\caption{Network scalability comparison with different numbers of MTS blocks ($NB$) on RGB images.
}
\vspace{-0.3cm}
\begin{center}
\begin{adjustbox}{width=0.5\textwidth,center}
\begin{tabular}{ c| ccc|ccc }
\hline
\hline
\multirow{2}{*}{CR}
&\multicolumn{3}{c|}{\thead{Urban100}} & \multicolumn{3}{c}{\thead{DIV2K}} \\

        &  \multicolumn{1}{c}{\thead{ (T=12, NB=1)}}
        &  \multicolumn{1}{c}{\thead{ (T=12, NB=2)}}
         &  \multicolumn{1}{c|}{\thead{ (T=12, NB=3)}}
         &  \multicolumn{1}{c}{\thead{(T=12, NB=1)}}
        &  \multicolumn{1}{c}{\thead{(T=12, NB=2)}}
         &  \multicolumn{1}{c}{\thead{ (T=12, NB=3)}} 
\\
\hline
\hline   
10\%&27.86/0.8615&28.41/0.8750&\textcolor{red}{\underline{28.53}}/\textcolor{red}{\underline{0.8773}}&32.33/0.9097&32.68/0.9189&\textcolor{red}{\underline{32.91}}/\textcolor{red}{\underline{0.9230}} \\
30\%&34.52/0.9580&35.39/0.9651&\textcolor{red}{\underline{36.50}}/\textcolor{red}{\underline{0.9709}}&38.70/0.9746&39.66/0.9789&\textcolor{red}{\underline{40.85}}/\textcolor{red}{\underline{0.9833}}\\
50\%&\textcolor{red}{\underline{45.31}}/\textcolor{red}{\underline{0.9940}}&43.46/0.9924&43.12/0.9918&\textcolor{red}{\underline{49.38}}/\textcolor{red}{\underline{0.9968}}&48.17/0.9959&47.79/0.9956 \\     
Avg. &35.90/0.9378&35.75/0.9442&\textcolor{red}{\underline{36.05}}/\textcolor{red}{\underline{0.9467}}&40.14/0.9604&40.17/0.9646&\textcolor{red}{\underline{40.52}}/\textcolor{red}{\underline{0.9673}} \\
\hline
\hline
\end{tabular}
\end{adjustbox}
\end{center}
\label{tab:netscal}
\end{table}

\subsection{Ablation Studies}

\textbf{Network Scalability:} 
We evaluate scalability by varying the number of MTS blocks ($NB$) and summation depth ($T$). As shown in Table~\ref{tab:netscal}, increasing the number of MTS blocks consistently improves reconstruction performance, confirming the positive impact of network depth.

\noindent \textbf{Activation Functions:} 
We compare ReLU, MHG~\cite{Xu2025MultiScaleTS}, and Identity activations in the CS encoder with $T=12, NB=3$. As reported in Table~\ref{tab:act}, MHG yields the best PSNR, while Identity achieves the highest SSIM on both Urban100 and DIV2K.

\begin{figure}[t]
\begin{center}
    \includegraphics[width=0.8\linewidth]{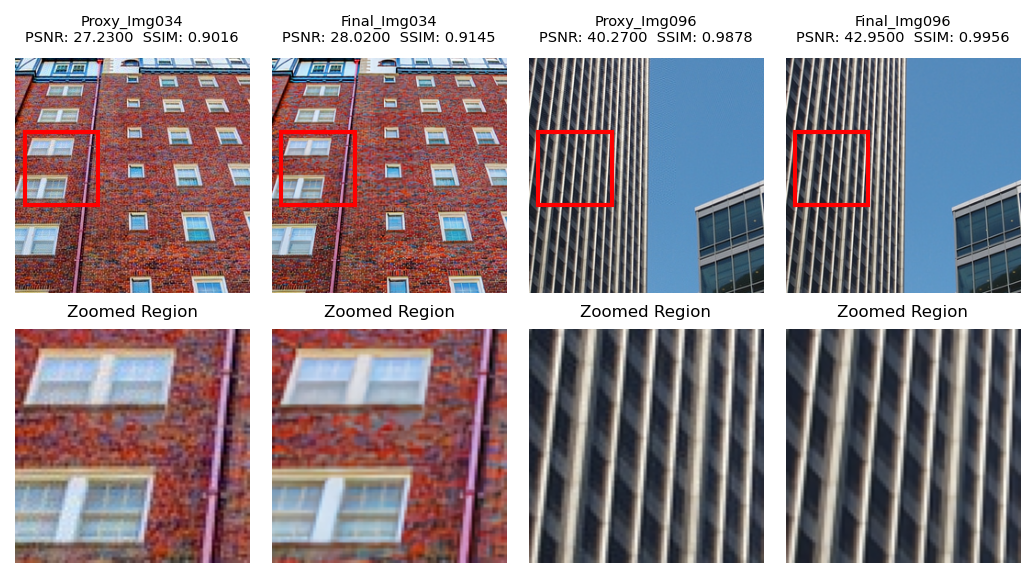}
\end{center}
   \vspace{-0.6cm}
   \caption{Comparison between proxy reconstructions (adjoint output) and final results produced by MTS-CSNet.
}
\label{fig:proxy}
\end{figure}

\noindent \textbf{Proximal Reconstruction:} 
Figure~\ref{fig:proxy} shows that the MTS-layer-based autoencoder produces high-quality proxy reconstructions. Compared to the final outputs, the proximal results incur only minor degradation (0.79/2.68 dB PSNR on img034 and img096, respectively), demonstrating the effectiveness of the learned adjoint.

\begin{table}[t]
\caption{Effect of different nonlinearities in the adjoint back-projection of the MTS-layer-based autoencoder, evaluated using PSNR (dB) at a sampling rate of 0.3 with $T=12$ and $NB=3$.}
\begin{adjustbox}{width=0.3\textwidth,center}
 \begin{tabular}{c|| c  c} 
 \hline
 \hline
 Activation & Urban100 & DIV2K \\  
 \hline
 MHG      & \textcolor{red}{\underline{36.50}}/\textcolor{red}{\underline{0.9709}} & \textcolor{red}{\underline{40.85}}/\textcolor{red}{\underline{0.9833}} \\ 
 Identity & 35.71/0.9670 & 40.23/0.9811 \\ 
 ReLU     & 29.06/0.8938 & 33.56/0.9358 \\ 
 \hline
 \hline
 \end{tabular}
 \end{adjustbox}
\label{tab:act}
\end{table}

\section{Conclusion}
\label{sec:con}

We presented MTS-CSNet, a non-iterative compressive sensing framework based on multiscale tensor summation. By using the MTS operator as a learnable sensing matrix, introducing a nonlinear adjoint back-projection, and refining reconstructions with a lightweight MTSNet, the proposed method operates entirely in tensor space with high efficiency. Experimental results demonstrate that MTS-CSNet achieves reconstruction quality comparable to or better than recent diffusion-based CS methods while requiring significantly lower computational cost and orders-of-magnitude faster inference. These results highlight the potential of multiscale tensor operators as an effective foundation for efficient compressive sensing of high-dimensional data.

\section*{Acknowledgment}
This work was supported by the Geometric AI Research to Business (R2B) project funded by Business Finland.



\bibliographystyle{IEEEtran}
\bibliography{refs}

\end{document}